\documentclass[sigconf]{acmart}
\AtBeginDocument{%
  }

\copyrightyear{2026}
\acmYear{2026}
\setcopyright{cc}
\setcctype{by}
\acmConference[FutureHCI '26]{The 2026 Future of HCI Workshop}{August 17--18, 2026}{Blacksburg, VA, USA}
\acmBooktitle{The 2026 Future of HCI Workshop (FutureHCI '26), August 17--18, 2026, Blacksburg, VA, USA}
\acmDOI{10.1145/3860405.3860626}
\acmISBN{979-8-4007-2732-0/2026/08}

\begin{document}

\title{Reflective Nature Experiences in Virtual Reality}

\author{Shiva Ghasemi}
\orcid{0000-0003-1435-6342}
\affiliation{%
  \institution{Virginia Tech}
  \city{Blacksburg}
  \state{Virginia}
  \country{USA}
}
\email{shivagh@vt.edu}

\author{Maddie Oehler}
\orcid{0009-0005-7199-334X}
\affiliation{%
  \institution{Virginia Tech}
  \city{Blacksburg}
  \state{Virginia}
  \country{USA}
}
\email{mmo815@vt.edu}

\author{Annie Xenakis}
\orcid{0009-0003-0501-8357}
\affiliation{%
  \institution{Virginia Tech}
  \city{Blacksburg}
  \state{Virginia}
  \country{USA}
}
\email{anniex@vt.edu}

\author{Jeffrey Jiarui Yang}
\orcid{0009-0000-2527-1075}
\affiliation{%
  \institution{Virginia Tech}
  \city{Blacksburg}
  \state{Virginia}
  \country{USA}
}
\email{jeffreyjr@vt.edu}

\author{Qiuya Chen}
\orcid{0009-0009-1434-6057}
\affiliation{%
  \institution{Virginia Tech}
  \city{Blacksburg}
  \state{Virginia}
  \country{USA}
}
\email{qiuya@vt.edu}

\author{Scott McCrickard}
\orcid{0000-0001-9839-7146}
\affiliation{%
  \institution{Virginia Tech}
  \city{Blacksburg}
  \state{Virginia}
  \country{USA}
}
\email{mccricks@cs.vt.edu}

\renewcommand{\shortauthors}{Ghasemi et al.}

\begin{abstract}
This paper presents the design and evaluation of an accessible virtual reality (VR) nature experience developed to promote mindfulness, reduce stress, and support emotional well-being. The system immerses users in a calming natural environment through realistic landscapes, ambient soundscapes, guided breathing exercises, and interactive checkpoints that encourage reflection. Accessibility was a central design consideration, particularly for users with diverse physical and sensory abilities or limited access to natural environments and traditional mindfulness resources. A pilot study with 10 participants indicated that the experience could support relaxation and emotional well-being. Participants also reported greater immersion with continuous movement than with teleportation, highlighting the importance of intuitive locomotion in designing inclusive, nature-based VR experiences for mindfulness and stress reduction.
\end{abstract}



\begin{teaserfigure}
    \centering
    \includegraphics[width=0.75\textwidth]{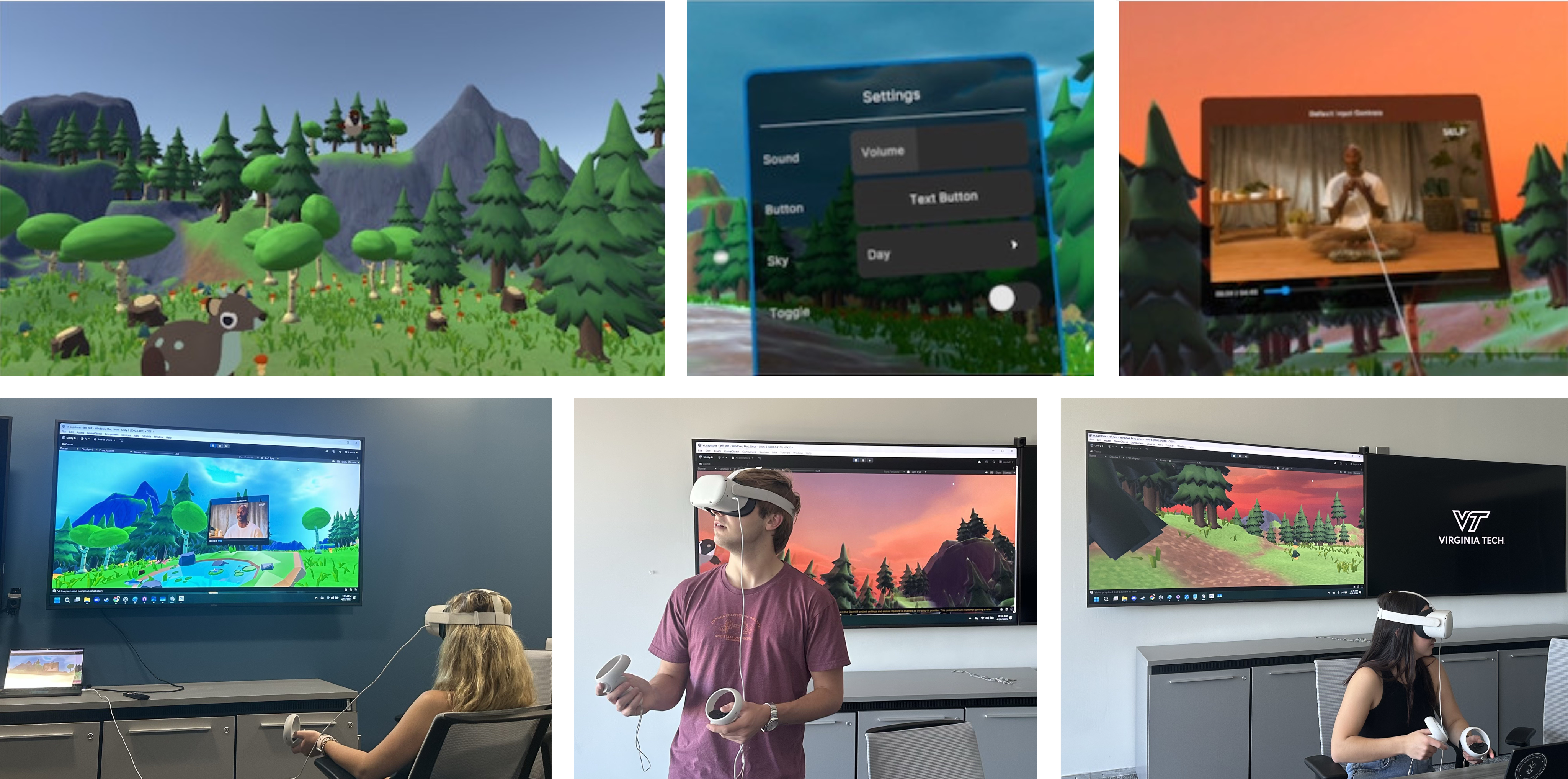}
    \caption{Screenshots of guided meditation, environmental interactions, and scenery in the VR nature experience.}
    \Description{A montage of screenshots from the VR nature experience showing guided meditation, immersive natural environments, interactive checkpoints, and user interactions throughout the virtual hiking experience.}
    \label{fig:teaser}
\end{teaserfigure}

\received{09 June 2026}
\received[revised]{July 2026}
\received[accepted]{July 2026}

\maketitle

\section{Introduction}
Mental health challenges have become a significant concern among university students globally. The World Health Organization reports that in 2019, approximately 970 million people worldwide were living with a mental disorder, with anxiety and depression among the most prevalent conditions~\cite{who_mental_health}. Within university populations, these issues are particularly pronounced, with studies reporting that 75\% of students experience symptoms of depression, 88.4\% experience anxiety, and 84.4\% experience stress~\cite{asif2020frequency}. Despite this high prevalence, traditional interventions such as counseling and pharmacological treatments often provide insufficient or delayed support due to cost and limited availability. These barriers highlight the need for accessible and scalable mental health tools tailored to students, leading to growing interest in alternative therapeutic technologies that can complement conventional care by promoting relaxation and emotional well-being.

Stress can impair academic performance by disrupting attention, memory, and decision-making, while increasing mental fatigue and reducing information retention. These effects may be partly linked to elevated cortisol levels that interfere with brain regions involved in learning. Although conventional approaches such as therapy can be effective, limitations in accessibility and immediacy motivate the development of technology-mediated methods for stress reduction~\cite{10.1145/3706599.3720197,10.1145/3613905.3651132}.

Virtual reality (VR) is a technology that immerses users in computer-generated three-dimensional environments, fostering a sense of presence and enabling individuals to experience the simulated setting as though they are physically situated within it. VR delivers a multi-sensory experience that includes visual, auditory, tactile, and even olfactory inputs. The user’s perception of truly inhabiting a virtual environment is referred to as presence, often described as the subjective sense of “being there” within a mediated environment~\cite{williams2025charting}. Since the 1990s, VR technologies have seen a steady increase in application for the prevention and treatment of various psychological and physical health conditions~\cite{white_prescription_2018}. VR has emerged as a particularly promising therapeutic tool, especially within the domain of mental health interventions, due to its unique capacity to facilitate the creation of immersive, controlled, and highly customizable therapeutic environments. 

The versatility of this medium is demonstrated by its application across a broad spectrum of clinical needs. Specifically, VR has been effectively integrated into the treatment of anxiety disorders and phobiass~\cite{maples2017use}, post-traumatic stress disorder~\cite{kothgassner2019virtual}, and obsessive-compulsive disorder. Beyond these, researchers have explored its utility in depression and stress management, substance use disorders~\cite{bisso2020immersive}, and schizophrenia~\cite{bisso2020immersive}, Furthermore, the technology has proven valuable in addressing eating disorders~\cite{ferrer2012use} and rehabilitating cognitive impairments resulting from stroke or brain injury~\cite{gamito2017cognitive}.


\section{Related Work}
The intersection of immersive technology and mental well-being has produced a growing body of research~\cite{ spangenberger_becoming_2022, bjorling_exploring_2022}, which can be broadly grouped into three areas: benefits of nature exposure, effectiveness of virtual nature interventions, and the use of VR in mindfulness and meditation practices. Research consistently demonstrates that even brief exposure to natural settings, such as forests, parks, and coastal areas, yields significant health benefits, particularly for urban residents~\cite{gascon2015mental}. These "green" and "blue" spaces serve as vital resources for psychological restoration. Webber et al.~\cite{webber_hci_2023} further conceptualize these interactions by categorizing nature engagement into mediated, abstracted, and simulated forms. This framework positions VR as a "simulated" interaction that provides a degree of separation from the physical world while offering a powerful, immersive representation of nature designed to support mental health when physical access is restricted. 

Our prior research shows that sharing nature and outdoor experiences can benefit people unable to engage in the experience themselves \cite{WangWritingHome2025,McCrickardHCIOutdoors2020}. This suggests promise for mental health effects from virtual representations of nature and the outdoors.

Building on traditional nature-based therapies, emerging research highlights VR as an effective medium for stress reduction and emotional regulation. Studies indicate that VR nature exposure is significantly more effective at alleviating stress than urban virtual environments~\cite{chan_nature_2023}. These interventions have shown utility in diverse contexts, ranging from clinical distraction during painful treatments to rehabilitation for neurological disorders~\cite{white_prescription_2018}. Innovative approaches, such as embodying a tree to increase "nature relatedness," suggest that higher levels of immersion directly correlate with stronger perceived connections to the environment~\cite{spangenberger_becoming_2022}. Furthermore, short-term virtual exposure has been shown to improve mood, attention, and relaxation across various demographics, including adolescents, who report high engagement with self-administered VR stress management tools~\cite{owens_effect_2023, bjorling_exploring_2022}.

\begin{figure}[ht]
    \centering
    \includegraphics[width=\columnwidth]{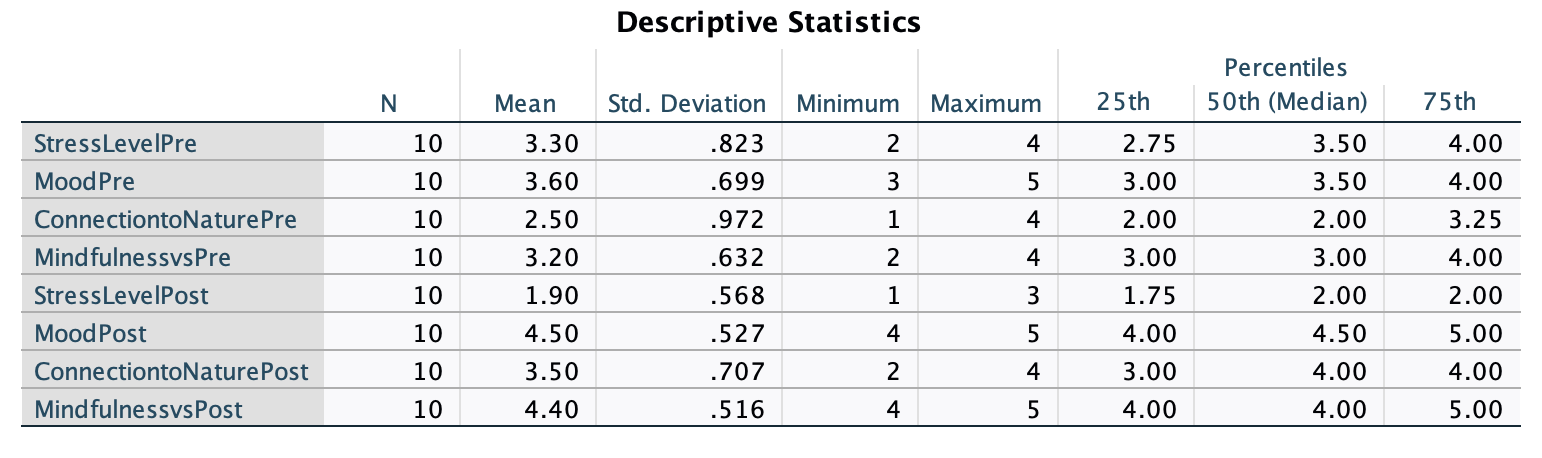}
    \caption{Descriptive statistics of pre- and post-intervention stress, mood, nature connectedness, and mindfulness.}
    \label{fig:descriptive_statistics}
\end{figure}

\begin{figure}[ht]
    \centering
    \includegraphics[width=\columnwidth]{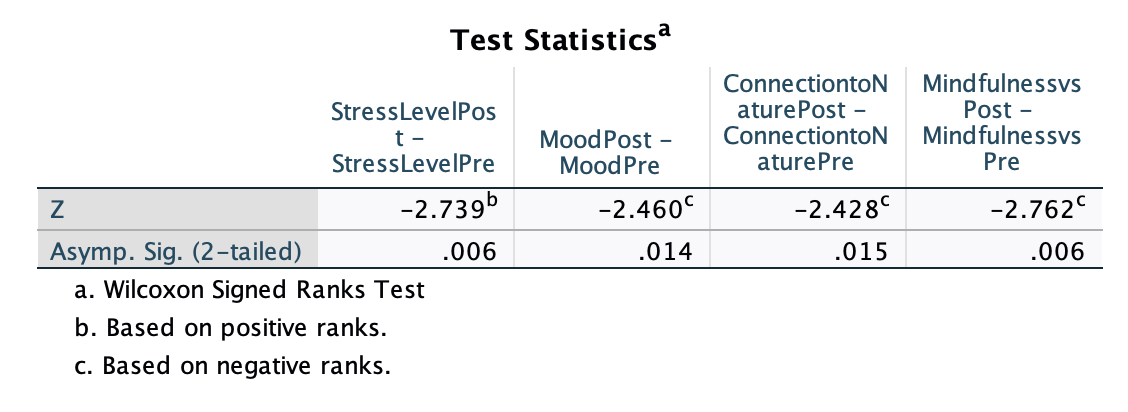}
    \caption{Wilcoxon signed ranks test results comparing pre- and post-intervention scores.} 
    \label{fig:test_statistics}
\end{figure}

VR is increasingly utilized to facilitate mindfulness, defined as non-judgmental focus on the present moment~\cite{bartlett2021mindfulness}. Research suggests that VR environments can outperform traditional audio-based mindfulness practices, particularly in promoting "decentering," which allows individuals to observe thoughts without emotional entanglement~\cite{chandrasiri_virtual_2020}. By combining tranquil visual stimuli, such as river landscapes, with natural ambient sounds and generative audio that responds to a user's breathing, VR creates a multisensory feedback loop that deepens awareness and presence~\cite{10.1145/3196709.3196765}. Design elements like gamification, auditory cues, and interactive features are critical to this success, as they help overcome barriers to traditional practice and provide customizable therapeutic experiences for individuals with limited mobility or restricted access to nature~\cite{li2025tuning}.

\section{Experiment}
This study examines the efficacy of immersive natural environments in facilitating stress reduction, mindfulness, and nature connectedness for populations with restricted access to outdoor spaces. We address RQ1: Which design features within a VR nature environment most effectively promote mindfulness and mitigate stress for individuals with limited access to natural settings or traditional mindfulness interventions?. We test the null hypothesis ($H_0$) that VR nature experience yields no significant effect on emotional well-being, against the alternative hypothesis ($H_1$) that engagement leads to measurable improvements in mindfulness and self-reported stress reduction relative to baseline. Adopting a mixed-methods approach, we triangulate quantitative psychological metrics with qualitative thematic analysis of user experiences. Data from surveys, open-ended responses, and observations were synthesized to identify recurring emotional and physiological patterns.

\subsection{Apparatus}
\label{Apparatus}
Participants used a Meta Quest 2 head-mounted display with an 89° horizontal and 93° vertical field of view (FOV), $1832 \times $1920 resolution per eye, and camera-based hand tracking.

\subsection{Design}
A simulated peaceful natural hiking experience that supports mindfulness, exploration, and relaxation has been designed, as shown in Figure~\ref{fig:sessions}. The scene features a winding mountain trail surrounded by diverse natural elements, including colorful landscapes, a serene pond, mushrooms, roaming deer, and lush vegetation. To enhance immersion, an animated bird companion guides users throughout the experience, providing a dynamic and engaging presence.
Users can select between two sky settings, a bright daytime sky or a vivid sunset, allowing for mood personalization. Three interactive checkpoints are integrated along the trail, where users engage in structured mindfulness activities such as guided breathing exercises and observational reflection of wildlife. Although the path provides subtle directional cues toward the mountain summit, the environment supports free exploration, enabling users to deviate from the main trail and interact with the surroundings.

\subsection{Tasks}
A standardized task sequence was used to evaluate the system’s design and psychological impact. The protocol began with an orientation phase where participants learned the interface, practiced navigation, and adjusted settings such as ambient audio and daytime or sunset mode. They then completed environmental observation tasks followed by a guided trail walk with checkpoints at a pond, meadow, and summit. Near the summit, participants performed a guided breathing exercise to encourage mindfulness and reflection.

\subsection{Participants}
The study included 10 participants (18–24 years; 7 female, 3 male) recruited from a university population. Eligible participants were 18 years or older, able to provide informed consent, and capable of safely using VR; individuals with medical conditions contraindicating VR use (e.g., epilepsy or severe motion sickness) were excluded. Most participants had prior VR experience (7 occasional, 2 once, 1 none), and 8 had previous mindfulness or meditation experience.

\section{Findings}
This section presents a mixed-methods analysis of the VR nature experience, integrating quantitative survey data with qualitative thematic findings. Pre- and post-intervention comparisons examine changes in stress, mood, mindfulness, and nature connectedness, while participants’ reflections provide deeper insight into the intervention’s emotional effects and overall user experience.

\begin{figure}[ht]
    \centering
    \includegraphics[width=\columnwidth]{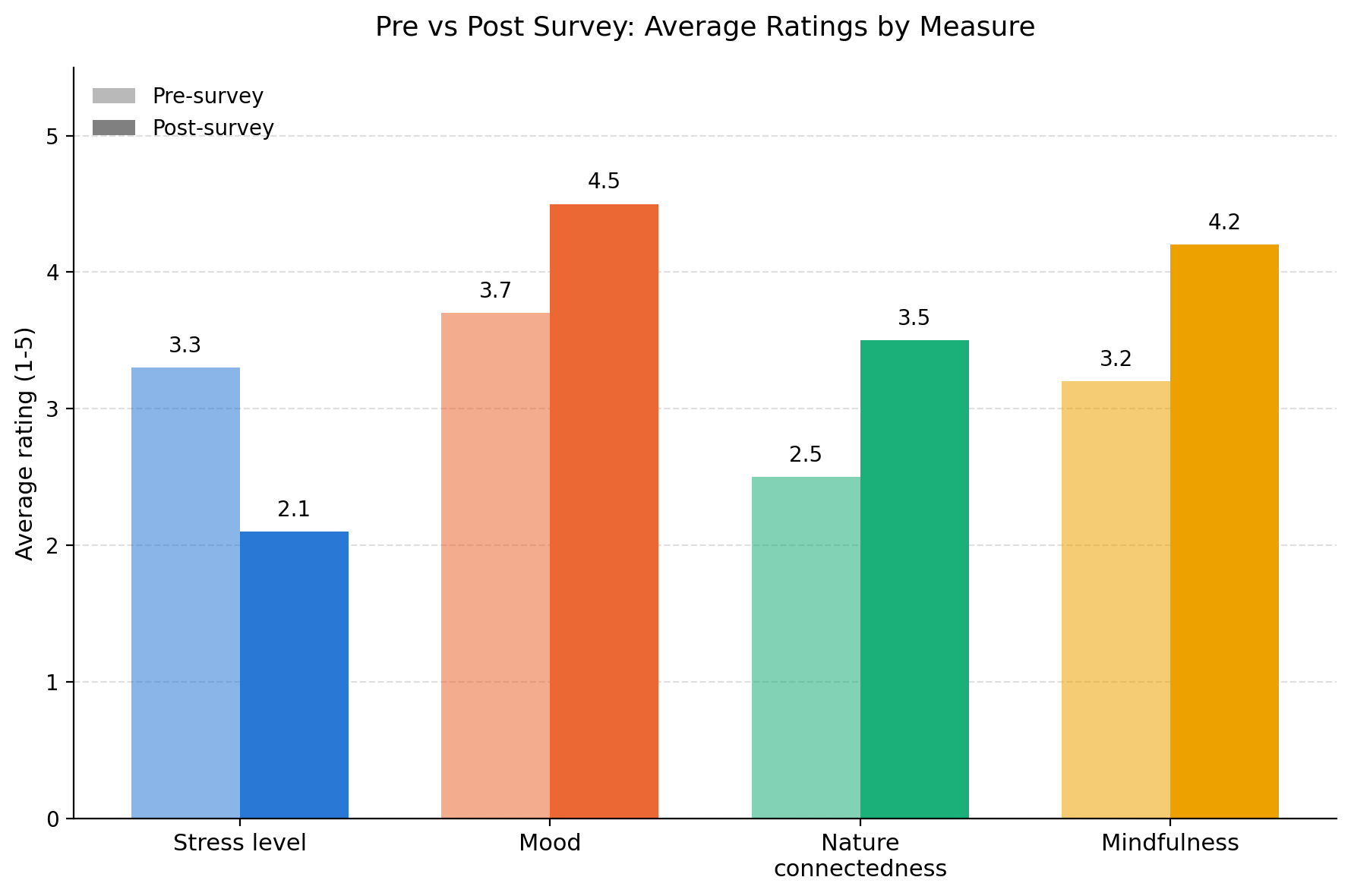}
    \caption{Bar graph of pre- and post-intervention stress, mood, nature connectedness, and mindfulness.}
    \label{fig:survey}
\end{figure}

\subsection{Pre and post Surveys}
Participants completed pre- and post-intervention surveys using 5-point Likert scales (1 = least, 5 = most) to measure changes in psychological well-being. The survey assessed four key dimensions: perceived stress, mood, connection to nature, and mindfulness. Descriptive statistics are presented in Figures~\ref{fig:descriptive_statistics} and~\ref{fig:test_statistics}, and survey data are visualized in Figure~\ref{fig:survey}.\vspace{0.6em}

\noindent \textbf{Stress Level}

\noindent Participants' mean stress levels decreased from $3.30$ ($SD = 0.82$) pre-intervention to $1.90$ ($SD = 0.57$) post-intervention, while mood ratings improved from $3.7$ to $4.5$. Additionally, the subjective sense of connection to nature rose from a mean of $2.5$ to $3.5$ following the VR experience. Given that the ordinal data were not normally distributed, a Wilcoxon signed-rank test was employed to assess the intervention's impact. The results confirmed a robust reduction in stress ($Z = -2.739, p = .006$), with $90\%$ of the cohort reporting lower stress levels and $10\%$ remaining unchanged. \vspace{0.6em}


\noindent \textbf{Mood}

\noindent Mood demonstrated a notable enhancement, aligning with findings from a separate study~\cite{spano2023virtual} that highlighted the positive influence of VR on emotional states. The mean mood score increased from 3.60 (SD = 0.70) to 4.50 (SD = 0.53), and the Wilcoxon test result was statistically significant (Z = -2.460, p = .014). Seven participants reported improved mood, while three showed no change. \vspace{0.6em}
 
\noindent \textbf{Nature Connectedness}

\noindent A significant increase was also observed in participants’ connection to nature. The mean score rose from 2.50 (SD = 0.97) to 3.50 (SD = 0.71), with the Wilcoxon test showing statistical significance (Z = -2.428, p = .015). Seven participants exhibited increased connectedness to nature, and three remained unchanged. \vspace{0.6em}

\noindent \textbf{Mindfulness}

\noindent Mindfulness scores increased significantly post-intervention. The mean score rose from 3.20 (SD = 0.63) to 4.40 (SD = 0.52), with the Wilcoxon test yielding a significant result (Z = -2.762, p = .006). Nine participants reported higher mindfulness scores post-intervention, with one reporting no change. 

\subsection{Thematic Analysis}
Thematic analysis is a qualitative method for systematically identifying, organizing, and interpreting patterns within unstructured data~\cite{lazar2017research}. In this study, it was used to analyze participants’ post-experience reflections and observational feedback collected during the VR nature experience. All qualitative data were transcribed and independently open-coded by four researchers to generate initial codes and categories~\cite{saldana2021coding}. The researchers then compared their codes and resolved discrepancies through discussion until consensus was reached. Axial coding was subsequently conducted to examine relationships among the codes and derive higher-level themes. The resulting themes and representative participant quotations are presented in Table~\ref{tab:condensed-quotes}.
\vspace{0.6em}

\noindent \textbf{\textit{Sense of Escape and Psychological Distance from Stressors}}

\noindent Participants consistently reported reduced anxiety and increased tranquility, describing the VR experience as a psychological escape from daily stressors. The environment supported emotional decompression and cognitive relaxation by enabling mental disengagement from intrusive thoughts and external pressures. A strong sense of spatial presence contributed to this effect, with participants noting they \textit{“forgot [they] were in a lab setting.”} Naturalistic visuals and ambient soundscapes further reinforced psychological distancing, enabling sustained attention and mindfulness. Participant feedback, such as \textit{“I felt like I was away from my daily stress,”} highlights immersive design as a key mechanism for fostering a restorative mental state.
\vspace{0.6em}

\noindent \textbf{\textit{Sensory Congruence and Embodied Relaxation}}

\noindent Spatialized ambient audio, including avian vocalizations, footsteps, and dynamic lake sounds, enhanced realism and immersion. Most participants preferred continuous movement over teleportation, noting that auditory feedback from footsteps supported a more natural experience. Walking speed was calibrated to approximately 1.5 m/s (4.7–5.4 km/h), matching average human walking pace. One participant noted, \textit{“Hearing my footsteps as I walked made it feel like I was really there.”} These findings highlight the role of multisensory congruence in strengthening spatial presence and supporting immersion in VR environments.\vspace{0.6em}

\begin{table*}[ht]
\centering
\small
\caption{Thematic Analysis of Participant Experiences}
\label{tab:condensed-quotes}
\begin{tabular}{|p{4.5cm}|p{11.5cm}|}
\hline
\textbf{Theme} & \textbf{Key Participant Insights} \\
\hline
\textbf{Sense of Escape and Psychological Distance from Stressors} & ``Away from daily stress'' (P1); ``Clear my mind'' (P3); ``Relaxing'' (P2); ``Teleported to the woods'' (P8); ``A break from everything'' (P10). \\
\hline
\textbf{Sensory Congruence and Embodied Relaxation} & ``Footsteps made it feel real'' (P3); ``Birds/footsteps make you feel in the world'' (P6); ``Deer faded as I moved'' (P7); ``Water got louder'' (P9); ``Natural sound shifts'' (P10). \\
\hline
\textbf{Maintaining Visual Stability to Support Immersion and Comfort} & ``Perspective jumping is unnatural'' (P3); ``Change of perspective not fluid'' (P7); ``Sudden movement pulled me out'' (P6); ``Felt slightly dizzy'' (P9); ``Smooth transitions help relaxation'' (P8). \\
\hline
\textbf{Personal Preferences in VR Experience} & ``Adjust things based on feeling'' (P6); ``Easy to adjust volume'' (P4); ``Personalize the birds'' (P2); ``Choose sunset to match mood'' (P9); ``Control the pace and breathe'' (P10). \\
\hline
\textbf{Moments of Discovery in Virtual Nature Experience} & ``Like a 3D Google Map'' (P1); ``Surprised by deers'' (P7); ``Cool to have a buddy'' (P3); ``Glowing mushrooms were magical'' (P8); ``Smiled at the fish'' (P9). \\
\hline
\textbf{Emotional Connectedness to Nature} & ``Reminded me nature is beautiful'' (P5); ``Atmosphere reels you in'' (P4); ``Loved the sun setting'' (P5); ``Want to go hiking in real life'' (P8); ``Calmer and closer to nature'' (P10). \\
\hline
\textbf{Limitations of VR Hiking in Replicating Physical Hiking Experience} & ``Missed moving my whole body'' (P4); ``Preferred real hiking'' (P7); ``Not a replacement for physical activity'' (P6); ``Terrain felt too smooth/no challenge'' (P9); ``Missed feeling the wind or sun'' (P10). \\
\hline
\end{tabular}
\end{table*}

\noindent \textbf{\textit{Maintaining Visual Stability to Support Immersion}}

\noindent Participants emphasized that visual stability and a continuous viewpoint are essential for sustaining immersion and emotional regulation in VR experience. Sudden perspective shifts were described as disruptive, reducing presence and causing mild disorientation. This effect may be influenced by hardware limitations such as the Meta Quest 2 FOV, which is narrower than human peripheral vision and can affect spatial awareness. Locomotion design also impacted stability, with teleportation perceived as abrupt and unnatural, while continuous movement aligned with natural walking pace was preferred for smoother transitions and realism. Participants noted that \textit{“Perspective control jumping is unnatural, and I prefer smoother transitions,”} and that \textit{“I felt most relaxed when I was in control of where I looked.”} These findings highlight the importance of stable visual flow and user-aligned movement design in supporting comfort and presence in VR nature experiences.

\vspace{0.6em}

\noindent \textbf{\textit{Variability and Personal Preferences in VR Experience}}

\noindent Participants emphasized customization as important for control, engagement, and emotional connection. Simple adjustments such as switching between daytime and sunset settings or modifying ambient audio shifted users from passive observers to active participants, supporting emotional regulation through increased agency. Flexible interaction modes also improved accessibility by accommodating different physical needs and levels of VR experience through seated and standing options with adjustable pacing. 
\vspace{0.6em}

\noindent \textbf{\textit{Moments of Discovery in Virtual Nature Experience}}

\noindent Participants reported positive emotional reactions, including excitement, amusement, and surprise, especially when encountering deer or coming across unexpected features within the virtual environment. Notably, a subset of participants expressed a desire for a higher degree of visual realism and geographic accuracy. 
For example, P1 remarked, \textit{“This experience could be like a 3D Google Map that simulates nature.”} This feedback reflects growing user expectations for VR environments to serve as highly detailed, interactive representations of real-world locations—an aspiration consistent with the concept of the “digital twin.” \vspace{0.6em}

\noindent \textbf{\textit{Emotional Connectedness to Nature}}

\noindent Several participants described experiencing an emotional connection to nature while engaging with the VR environment. The immersive natural scenery, wildlife encounters, and ambient sounds fostered feelings of appreciation and attachment to the natural world, even within the simulated setting. This sense of biophilic engagement contributed to the participants' overall relaxation and emotional well-being. 
\textit{"Even though it was virtual, it reminded me how peaceful and beautiful nature is — it made me want to go outside more."}\vspace{0.6em}

\noindent \textbf{\textit{Challenges in Replicating Physical Hiking}}

\noindent While the VR intervention provided immersive and restorative experiences, it was characterized by participants as a partial substitute for real-world hiking rather than a full equivalent. Traditional hiking integrates complex physiological demands, including cardiovascular exertion, muscular engagement, and the coordination required to navigate varied terrains. These embodied elements are instrumental in psychological well-being through the release of endorphins and the regulation of cortisol levels. Conversely, while VR hiking is visually and emotionally salient, it lacks the physical exertion and comprehensive multi-sensory integration of authentic outdoor movement. As one participant noted, \textit{“It was relaxing, but I missed the feeling of moving my whole body — part of hiking for me is the workout and how good I feel afterward.”} 

\section{Discussion}

This pilot study demonstrates that immersive virtual nature experiences can positively influence users' perceived stress, mood, mindfulness, and connection to nature. Quantitative data revealed statistically significant improvements across all measured domains: participants reported reduced stress, enhanced mood, increased mindfulness, and a stronger sense of connection to nature. These results support the alternative hypothesis ($H_{1}$) and align with prior research~\cite{white_prescription_2018,chan_nature_2023} highlighting the restorative potential of virtual natural environments. On average, stress ratings decreased by 1–2 points on a 5-point scale, suggesting that the immersive experience supported emotional regulation. 

The qualitative data provide further context for these outcomes, illustrating a consistent reduction in stress, improvement in mood, and a heightened sense of connectedness to nature. These results suggest that immersive virtual environments can partially replicate the psychological benefits associated with real-world nature exposure. Participants attributed these effects to intentional design elements, including high-fidelity ambient soundscapes, naturalistic visuals, and guided breathing exercises. The inclusion of low-interaction accessibility features such as seated and standing modes was also important in maintaining immersion and comfort, supporting deeper emotional engagement with the therapeutic experience.

Thematic analysis revealed three key mechanisms underlying the observed effects. First, restorative escapism provided participants with a temporary psychological refuge that supported cognitive relief from daily stressors, reflecting a form of therapeutic disengagement that enables emotional regulation and attentional reset. Second, sensory and locomotion congruence was central, as participants preferred continuous, naturalistic movement over teleportation, which enhanced immersion and aligned with embodied cognition principles. Third, user agency through customization features such as adjustable sky settings, ambient audio, and exploration freedom fostered control and personal relevance, supporting diverse emotional needs. Together, these mechanisms strengthened presence and psychological distancing from stressors. Guided breathing exercises and ambient cues supported mindfulness by directing attention toward present-moment awareness.

Despite these positive outcomes, some usability limitations were observed. A few participants initially experienced minor difficulties with navigation and interface controls, particularly due to inexperience with VR systems rather than design flaws. Navigation ambiguity, especially near the trail summit, suggests that clearer environmental cues could improve usability. Minor motion discomfort and visual fidelity concerns were also reported, though no significant motion sickness occurred. These findings indicate opportunities for refinement in locomotion design, environmental guidance, and graphical consistency.

While VR nature experience does not replicate the full physiological benefits of real-world nature exposure, such as cardiovascular exertion or multisensory inputs like wind and temperature, it provides meaningful psychological benefits. As hardware advances, particularly in field of view, refresh rate, and motion tracking, VR experiences are becoming increasingly immersive. When combined with biometric feedback systems that respond to stress or engagement in real time, VR has the potential to evolve into a more adaptive and personalized therapeutic medium. Importantly, VR nature experience should be viewed as a complementary, not substitutive, intervention. Its greatest value lies in expanding access to nature-inspired mental health support for individuals who face barriers due to mobility, geography, or time constraints. This positions VR as a scalable and inclusive tool for use in clinical settings, educational programs, workplace wellness initiatives, and elder care environments.

This paper makes three primary contributions to the design of restorative immersive experiences. First, we present the design and implementation of an accessible and interactive VR nature experience that integrates guided mindfulness, environmental exploration, user personalization, and continuous locomotion to support relaxation and emotional well-being. Second, we provide a mixed-methods evaluation demonstrating significant improvements in perceived stress, mood, mindfulness, and nature connectedness while revealing the interaction mechanisms that contributed to these outcomes through quantitative and qualitative analyses. Finally, we derive a set of HCI design implications for future restorative immersive systems, highlighting the roles of continuous locomotion, multisensory congruence, environmental personalization, guided reflection, and accessible interaction design in creating engaging and psychologically supportive immersive experiences.

\section{Future Work and Impact}
To build upon the initial insights gained from this pilot evaluation and directly align our trajectory with emerging paradigms in HCI, we outline a two-fold research roadmap encompassing short-term technical enhancements and a long-term vision for pervasive, accessible interaction. In the near term, future work will focus on transitioning the virtual environment from manual customization to dynamic, real-time adaptation. To reduce the cognitive burden of adjusting settings, the system will use multimodal sensing, including head-mounted eye tracking, to monitor indicators such as blink rate, saccades, pupil dilation, gaze fixation, and dwell time. These signals can support a closed-loop system that automatically adjusts sound, lighting, and breathing guidance in response to changes in stress, cognitive load, and visual attention. Over the next five years, this research aims to extend restorative immersive experiences beyond stationary VR into lightweight, context-aware AR and XR systems. Such systems could deliver brief, personalized visual and auditory interventions in everyday settings. Future studies will also involve collaboration with clinical psychology and occupational therapy experts to validate safety and therapeutic effectiveness. Integrating geospatial data and digital twins of natural environments may further provide realistic and accessible restorative experiences for individuals facing mobility, geographic, or financial barriers.


\section{Conclusion}
This study demonstrates the potential of VR nature experiences as accessible interventions for enhancing emotional well-being, particularly for individuals facing barriers to natural environments or traditional mindfulness practices. By providing a "psychological escape" through immersive landscapes, the system supports effective stress management and emotional regulation. Ultimately, these findings validate the role of VR in facilitating nature-inspired, mindful interactions. 

\bibliographystyle{ACM-Reference-Format}
\bibliography{sample-base}

\end{document}